\documentclass[11pt]{article}

\usepackage{cite}
\usepackage{color}

\usepackage{graphicx}
\usepackage{subcaption} % For subfigure

\usepackage{amsmath}
\usepackage{dsfont} % For \mathds{K}
\usepackage{siunitx}

\usepackage{tikz}
\usetikzlibrary{arrows} % For arrow width
\usetikzlibrary{calc} % For coordinate sum
\usetikzlibrary{through} % For circle through
\usepackage{pgfplots}

\usepackage{easyReview}

\begin{document}

\title{Trajectory Generation using Sharpness Continuous Dubins-like Paths with Applications in Control of Heavy-Duty Vehicles\footnote{This work was partially supported by the Wallenberg AI, Autonomous Systems and Software Program (WASP) funded by the Knut and Alice Wallenberg Foundation}}

\author{Rui Oliveira$^{1, 2}$, Pedro F. Lima$^{1, 2}$, Marcello Cirillo$^{2}$, \\Jonas M{\aa}rtensson$^{1}$ and Bo Wahlberg$^{1}$
\thanks{$^{1}$Department of Automatic Control, School of Electrical Engineering and Computer Science, KTH Royal Institute of Technology, Stockholm, Sweden
        {\tt\small rfoli@kth.se}, {\tt\small pfrdal@kth.se}, {\tt\small jonas1@kth.se}, {\tt\small bo@kth.se}}%
\thanks{$^{2}$Scania, Autonomous Transport Solutions, S{\"o}dert{\"a}lje, Sweden {\tt\small marcello.cirillo@scania.com} }%
}

\date{}

\begin{titlepage}
\maketitle
\thispagestyle{empty}

%%%%%%%%%%%%%%%%%%%%%%%%%%%%%%%%
\begin{abstract}
We present a trajectory generation framework for control of wheeled vehicles under steering actuator constraints.
The motivation is smooth driving of autonomous heavy-duty vehicles, which are characterized by slow actuator dynamics.
In order to deal with the slow dynamics, we take into account rate and, additionally, torque limitations of the steering actuator directly.
Previous methods only take into account limitations in the path curvature, which deals indirectly with steering rate limitations.
We propose the new concept of Sharpness Continuous curves, which uses cubic curvature paths together with circular arcs to steer the vehicle.
The obtained paths are characterized by a smooth and continuously differentiable steering angle profile.
The final trajectories computed with our method provide low-level controllers with reference signals which are easier to track, resulting in improved performance.
The smoothness of the obtained steering profiles also results in increased passenger comfort.
The method is characterized by fast computation times.
We detail possible path planning applications of the method, and conduct simulations that show its advantages and real-time capabilities.

\end{abstract}

\end{titlepage}

\def\maxcurvature{\kappa_{\max}} % This looks better than \text{max}

\def\maxsteering{\phi_{\max}}
\def\maxsteeringdot{\dot{\phi}_{\max}}
\def\maxsteeringdotdot{\ddot{\phi}_{\max}}

\def\peaksteeringdot{\dot{\phi}_{\mathrm{peak}}}
\def\peaksteeringdotdot{\ddot{\phi}_{\mathrm{peak}}}

% G-three path property
\def\gthree{\mathbf{G}^3}

% Corresponds to the start configuration
% Where the vehicle starts its movement
\def\startconfiguration{\mathbf{q}_{\mathrm{s}}}
% Corresponds to the goal configuration
% Where the vehicle tries to move to
\def\goalconfiguration{\mathbf{q}_{\mathrm{g}}}

% Corresponds to the tangent where the line segment starts
\def\TangentExitCfgSubscript{\mathrm{a}}
\def\TangentExitConfiguration{\mathbf{q}_{\TangentExitCfgSubscript}}
% Corresponds to the tangent where the line segment ends
\def\TangentEntryCfgSubscript{\mathrm{b}}
\def\TangentEntryConfiguration{\mathbf{q}_{\TangentEntryCfgSubscript}}

% For an SC Turn
% Corresponds to the start configuration
\def\ScTurnStartCfgSubscript{\mathrm{1}}
\def\ScTurnStartConfiguration{\mathbf{q}_{\ScTurnStartCfgSubscript}}
% Corresponds to the end configuration
\def\ScTurnEndCfgSubscript{\mathrm{4}}
\def\ScTurnEndConfiguration{\mathbf{q}_{\ScTurnEndCfgSubscript}}
% Corresponds to the arc start configuration
\def\ScTurnArcStartCfgSubscript{\mathrm{2}}
\def\ScTurnArcStartConfiguration{\mathbf{q}_{\ScTurnArcStartCfgSubscript}}
% Corresponds to the arc end configuration
\def\ScTurnArcEndCfgSubscript{\mathrm{3}}
\def\ScTurnArcEndConfiguration{\mathbf{q}_{\ScTurnArcEndCfgSubscript}}

\def\ScTurnCurvingPath{\Gamma_{\mathrm{1,2}}}
\def\ScTurnArcPath{\Gamma_{\mathrm{2,3}}}
\def\ScTurnDecurvingPath{\Gamma_{\mathrm{3,4}}}

% For connecting SC Turns
\def\StartCircleOriginalSubscript{\mathrm{s}}
\def\StartCircleOriginal{\Omega_{\StartCircleOriginalSubscript}}
\def\GoalCircleOriginalSubscript{\mathrm{f}}
\def\GoalCircleOriginal{\Omega_{\GoalCircleOriginalSubscript}}
\def\StartCircleAuxSubscript{\mathrm{a}}
\def\StartCircleAux{\Omega_{\StartCircleAuxSubscript}}
\def\GoalCircleAuxSubscript{\mathrm{b}}
\def\GoalCircleAux{\Omega_{\GoalCircleAuxSubscript}}

\def\tikzpicturesscale{0.5}
\def\tikzfontsscale{\normalsize}

\def\tikzmatlabfontsscale{\Large}

\def\AuxCircleDashPattern{dashed}
\def\AuxCircleRadiiDashPattern{densely dashed}

%%%%%%%%%%%%%%%%%%%%%%%%%%%%%%%%
\section{Introduction}
\subsection{Background and Motivation}

Path planning deals with the generation of paths or trajectories (paths with an associated time law) for a vehicle.
The generated paths/trajectories are used as a reference signal for the controllers implemented in the vehicle.
Planning methods for autonomous vehicles have come a long way from the initial problem of finding collision free paths, being now focused on properties such as kinodynamic constraints, optimality, and uncertainty~\cite{buehler2009darpa},~\cite{lavalle2006planning}.
Car-like vehicles must follow specific patterns of motion defined by their kinematic constraints.
These constraints introduce an additional difficulty, as they limit the maneuverability of the vehicle, resulting in limited types of paths that are admissible, \textit{i.e.}, can be feasibly followed.
Heavy-duty vehicles introduce additional constraints due to their slow actuator dynamics.
Thus planning methods for autonomous vehicles must be adapted, in order to deal with the additional control challenges imposed by heavy-duty vehicles. 

Much research effort has been devoted to the field of $\gthree$ path planning.
A $\gthree$ path is characterized by a continuously differentiable curvature profile.
$\gthree$ paths are important as they avoid jerky motion and wheel slippage~\cite{reuter1998}, simplifying the tracking task and improving controller performance~\cite{nagy2001trajectory}.
$\gthree$ path planning is analogous to the $\gthree$ interpolation problem with applications often related to Computer-Aided Design \cite{MCCRAE2009452}.
Some authors have focused on its applications to autonomous mobile robots~\cite{bianco2004smooth}.

Path planning for autonomous vehicles introduces additional demands.
Besides the $\gthree$ property, another important property of a path is its length.
Shorter paths are desirable as they result in more efficient driving.

Steering methods are a class of path planners that are able to efficiently compute a path between vehicle states in an environment without obstacles.
Even though the majority of autonomous vehicle applications considers obstacles, steering methods still prove to be useful, as they are often used as components of more complex path planners which can take obstacles into account~\cite{kuwata2009real}--\hspace{1sp}\cite{dolgov2010path}.

Dubins~\cite{dubins1957curves} and Reeds-Shepp~\cite{reeds1990optimal} paths are steering methods that connect two arbitrary vehicle poses through a minimal length path.
However, these paths have discontinuous curvatures, thus not being $\gthree$.
\cite{fraichard2004reeds} makes use of clothoidal paths to generate near length optimal curvature continuous paths.
Clothoids are an obvious choice, since they have long been used in road design, as they allow for smooth driving~\cite{lima2015clothoid}.
However, the curvature derivative is discontinuous, resulting in paths that are not $\gthree$.
\cite{parlangeli2010dubins} extends~\cite{dubins1957curves}, so that $\gthree$ paths with near optimal length are planned.
This last method deals with limitations of the path curvature and curvature derivative, which are not directly related to the steering actuator limitations of the vehicle.

\subsection{Main Contributions}

The contribution of this work comes from the generation of vehicle trajectories that:
\begin{itemize}
\item Directly take into account steering actuator magnitude, rate, and acceleration limitations, generating $\gthree$ paths;
\item Ease the controller task and improve passenger comfort;
\item Can connect arbitrary vehicle configurations;
\item Have fast computation times.
\end{itemize}

We build upon the work of~\cite{fraichard2004reeds}, replacing clothoids with cubic curvature paths.
Additionally, we formulate the constraints so that we can limit the actual steering angle rate and steering angle acceleration of the vehicle, instead of the curvature derivative of the path, as done in previous approaches~\cite{fraichard2004reeds},~\cite{parlangeli2010dubins}.
This is because the maximum steering angle rate and acceleration are more intuitive constraints that can be directly obtained from the vehicle actuator limitations.
The proposed method is also computationally fast and can be used online, as part of a more complex motion planner~\cite{videogames}.

\subsection{Outline}

Section \ref{sec:problem_statement} introduces the vehicle model used and defines the problem we address.
Section \ref{sec:sharpness_continuous_paths} presents Sharpness Continuous paths used to solve the stated problem.
Section \ref{sec:cubic_curvature_paths} presents Cubic Curvature paths, a building block of Sharpness Continuous paths.
Section \ref{sec:results} illustrates our simulation results, showing the performance of the method.
Conclusions and future work are presented in Section~\ref{sec:conclusions}.

%%%%%%%%%%%%%%%%%%%%%%%%%%%%%%%%
\section{Problem Statement}
\label{sec:problem_statement}

\subsection{Vehicle Model}

We start by defining the vehicle model as:
\begin{equation*} 
\dot{x} = v \cos \theta,
\qquad
\dot{y} = v \sin \theta,
\qquad
\dot{\theta} = v\kappa.
\end{equation*}
$(x, y)$ represents the location of the vehicle rear wheel axle center, $\theta$ its orientation and $v$ is the vehicle velocity.
The curvature $\kappa$ of a vehicle with wheelbase length $L$ is related to its steering angle $\phi$ through
\begin{align}
\kappa = \tan \left( \phi \right)/ L.
\label{eq:curvature_from_steering}
\end{align}
A vehicle pose is defined by the three variables $(x, y, \theta)$.
If an additional curvature is associated to a pose we obtain a configuration, defined as $(x, y, \theta, \kappa)$.

The steering angle of the vehicle is set by an actuator, which like any real system has physical limitations.
The limitations with which we comply in this work are:
\begin{itemize}
\item Maximum steering angle amplitude $\maxsteering$,
\item Maximum steering angle rate of change $\maxsteeringdot$,
\item Maximum steering actuation acceleration $\maxsteeringdotdot$.
\end{itemize}
These limitations effectively affect the vehicle motion capabilities and should be dealt with when generating paths.

\subsection{Path Feasibility}

Path feasibility depends on the capabilities of the vehicle that executes it and on the path itself.
The limited steering angle amplitude $\maxsteering$ imposes a maximum allowed curvature on the path $\maxcurvature$.
This limitation is addressed by generating paths which have a curvature profile $|\kappa| \leq \maxcurvature$~\cite{dubins1957curves, reeds1990optimal}.
Limited steering angle rate of change $\maxsteeringdot$ can be tackled by limiting the curvature derivative of the generated paths~\cite{fraichard2004reeds}.

In this paper, we deal with the third limitation, related to the limited steering angle acceleration $\maxsteeringdotdot$.
Having a limited $\maxsteeringdotdot$ results in $\dot{\phi}$ being a continuous function, which in turn indicates that $\phi$ is a continuously differentiable, $\mathbf{C^1}$ function.
The paths generated by~\cite{fraichard2004reeds} have corresponding $\dot{\phi}$ profiles with discontinuities, that require an infinite $\maxsteeringdotdot$.
This is impossible to achieve by an actuator, and motivates the usage of paths with a $\mathbf{C^1}$ steering profile.

The steering profile is related to the curvature profile through \eqref{eq:curvature_from_steering}.
The sharpness $\alpha$ is defined as the change of curvature along the path length $s$:
\begin{align*}
\alpha = \partial \kappa / \partial s.
\end{align*}
By ensuring sharpness continuity in a path, we guarantee that the curvature, and the steering profile of such a path is $\mathbf{C^1}$, \textit{i.e.}, the path is $\gthree$.
A vehicle is thus able to follow the path using a bounded steering acceleration $\maxsteeringdotdot$.

In the following section, we detail how to generate paths that respect all three limitations previously stated.

%%%%%%%%%%%%%%%%%%%%%%%%%%%%%%%%
\section{Sharpness Continuous Paths}
\label{sec:sharpness_continuous_paths}

In this section we present the Sharpness Continuous (SC) paths.
\ref{subsec:sc_paths_principle} introduces the principle behind SC paths.
SC paths are composed of SC turns (detailed in~\ref{subsec:sharpness_continuous_turns}) connected over a line segment.
The process of connecting SC turns over a line segment to form a continuous SC path is detailed in~\ref{subsec:connecting_sharpness_continuous_turns}.
When generating an SC path there is a total of 16 possible combinations of different SC turns that can be used.
\ref{subsec:sc_shortest_path} indicates how to choose the best combination.

\subsection{Principle}
\label{subsec:sc_paths_principle}

\cite{dubins1957curves, reeds1990optimal} use a combination of arc circle turns and/or line segments to connect two arbitrary poses.
In~\cite{fraichard2004reeds}, this idea is extended with Curvature Continuous (CC) turns, which replace arc circles by a combination of clothoid and arc circles.
We extend this further, replacing the clothoid segments by cubic curvature paths, achieving sharpness continuity and respecting the limited steering acceleration $\maxsteeringdotdot$.

\subsection{Sharpness Continuous Turns}
\label{subsec:sharpness_continuous_turns}

\def\tikzpicturesscale{0.65}
\begin{figure}
  \centering
     %\resizebox {0.8\columnwidth} {!} {
      %\input{tikz/sc_turn.tikz}
      %}
      \begin{tikzpicture}[scale=\tikzpicturesscale]

\newcommand*{\curvingcolor}{black!10!blue}
\newcommand*{\arccolor}{black!50!green}
\newcommand*{\decurvingcolor}{black!20!red}
\newcommand*{\pathtickness}{very thick}

\def\radiusinnercircle{4}

// The configuration entering the arc segment
\def\xqentry{ 3 }
\def\yqentry{ 1 }
\def\oqentry{ 30 }

// The orientation of the configuration exiting the arc segment
\def\oqexit{ 45 + 90 }

// The angle of the final pose
\def\oend{ 180 - 35 }

\path[ draw=black ,solid,line width=0.5mm,fill=black, preaction={-triangle 90,thin,draw,shorten >=-1mm} ] (0, 0) -- node[below] {\tikzfontsscale $\ScTurnStartConfiguration$ } ($(0, 0) + (1.5, 0)$);

\draw[\pathtickness, \curvingcolor] plot [smooth, tension=1] coordinates { (0,0) (1.5, 0.25) (\xqentry, \yqentry) } ;

\node [\curvingcolor, above] at (1.5, 0.25) {\hspace{-0.9em} \tikzfontsscale $\ScTurnCurvingPath$};

\def\circleinnercenterx{ {\xqentry + \radiusinnercircle*cos( \oqentry + 90 ) } }
\def\circleinnercentery{ {\yqentry + \radiusinnercircle*sin( \oqentry + 90 ) } }

\coordinate (circleinnercenter) at ( \circleinnercenterx , \circleinnercentery );

\draw [\AuxCircleDashPattern] (circleinnercenter) circle (\radiusinnercircle);

% Arc path
\draw[\pathtickness, \arccolor] (\xqentry, \yqentry) arc (\oqentry - 90:\oqexit - 90:\radiusinnercircle) ;
\node [\arccolor, right] at ({\circleinnercenterx + \radiusinnercircle} , \circleinnercentery) {\tikzfontsscale $\ScTurnArcPath$};

\coordinate (qexit) at ( {\circleinnercenterx + \radiusinnercircle*cos( \oqexit - 90 ) }  , {\circleinnercentery + \radiusinnercircle*sin( \oqexit - 90 ) }  );

\coordinate (qexitone) at ($(qexit) + (-1.5, 1.25)$);
\coordinate (qexittwo) at ($(\circleinnercenterx, {\circleinnercentery + \radiusinnercircle*sin( \oqexit - 90 )} ) + (0, 2)$);

%\draw[\pathtickness, \decurvingcolor] plot [smooth, tension=1] coordinates { (qexittwo) (qexitone) (qexit) };
\draw[\pathtickness, \decurvingcolor] plot [smooth, tension=1] coordinates { (qexit) (qexitone) (qexittwo) };

% q_i
\draw[\AuxCircleRadiiDashPattern] (circleinnercenter) -- (\xqentry, \yqentry) node[midway, right]{\tikzfontsscale $\maxcurvature^{-1}$};
\node [right] at (\xqentry, \yqentry) {\tikzfontsscale $\ScTurnArcStartConfiguration$} ;

% q_j
\draw[\AuxCircleRadiiDashPattern] (circleinnercenter) -- (qexit);
\node [right] at (qexit) {\tikzfontsscale $\ScTurnArcEndConfiguration$};

\node [\decurvingcolor, left] at (qexitone) {\tikzfontsscale $\ScTurnDecurvingPath$};

\node [draw, \AuxCircleDashPattern] at (circleinnercenter) [circle through={(qexittwo)}] {};

\node[circle,fill=black,inner sep=0pt,minimum size=4pt,label=left:{\tikzfontsscale $\Omega$}] (a) at (circleinnercenter) {};

\draw[\AuxCircleRadiiDashPattern] (circleinnercenter) -- (qexittwo) node[midway, left]{\tikzfontsscale $r$};

\path[ draw=black ,solid,line width=0.5mm,fill=black, preaction={-triangle 90,thin,draw,shorten >=-1mm} ] (qexittwo) -- node[midway, right] {\tikzfontsscale $\ScTurnEndConfiguration$ } ($(qexittwo) + ({1.5*cos( \oend )}, {1.5*sin( \oend )})$);

\draw[dashed] ($(qexittwo) + (-2,0)$) -- ($(qexittwo) + (2,0)$);

% Arc mu
\draw[->] ($(qexittwo) + ({1.0*cos( \oend )}, {1.0*sin( \oend )})$) arc (abs(\oend):180:1.0) node[midway, left]{\tikzfontsscale $\mu$};

% Arc sigma
\draw[->] ($(qexittwo) + ({1.0*cos( 0 )}, {1.0*sin( 0 )})$) arc (0:abs(\oend):1.0) node[midway, above]{\tikzfontsscale $\delta$};

\end{tikzpicture}
  \caption{Sharpness continuous turn general case. \label{fig:sharpness_continuous_turn} \vspace*{-3mm}}
\end{figure}
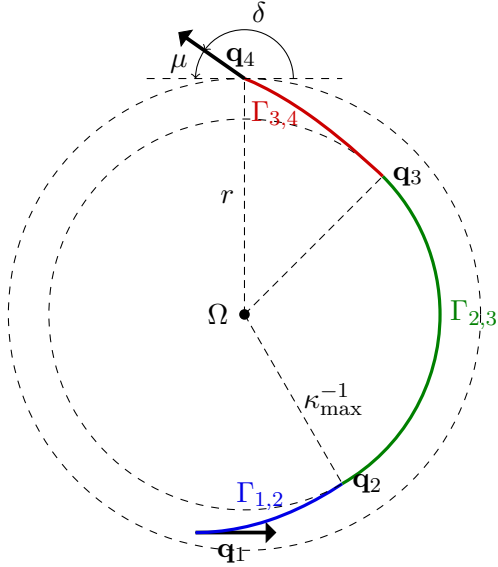

We propose Sharpness Continuous (SC) turns, which consist of three segments, an initial cubic curvature path $\ScTurnCurvingPath$, a circular arc $\ScTurnArcPath$, and a final cubic curvature path $\ScTurnDecurvingPath$.
Figure \ref{fig:sharpness_continuous_turn} shows an example of an SC turn.
The initial segment $\ScTurnCurvingPath$ starts at a configuration $\ScTurnStartConfiguration = (x_{\ScTurnStartCfgSubscript}, y_{\ScTurnStartCfgSubscript}, \theta_{\ScTurnStartCfgSubscript}, \kappa_{\ScTurnStartCfgSubscript})$ and ends with maximum curvature, $\pm \maxcurvature$, at a configuration $\ScTurnArcStartConfiguration = (x_{\ScTurnArcStartCfgSubscript}, y_{\ScTurnArcStartCfgSubscript}, \theta_{\ScTurnArcStartCfgSubscript}, \kappa_{\ScTurnArcStartCfgSubscript})$.
The second segment is a circular arc $\ScTurnArcPath$ with radius $\maxcurvature ^ {-1}$ and arbitrary arc length, starting at $\ScTurnArcStartConfiguration$ and ending at $\ScTurnArcEndConfiguration = (x_{\ScTurnArcEndCfgSubscript}, y_{\ScTurnArcEndCfgSubscript}, \theta_{\ScTurnArcEndCfgSubscript}, \kappa_{\ScTurnArcEndCfgSubscript})$.
The SC turn is completed with a path $\ScTurnDecurvingPath$, with starting curvature  $\pm \maxcurvature$ and ending at a configuration $\ScTurnEndConfiguration = (x_{\ScTurnEndCfgSubscript}, y_{\ScTurnEndCfgSubscript}, \theta_{\ScTurnEndCfgSubscript}, \kappa_{\ScTurnEndCfgSubscript})$.

We assume, without loss of generality, that the vehicle, and subsequently the path, starts at a configuration $\ScTurnStartConfiguration = (0, 0, 0, 0)$.
From $\ScTurnStartConfiguration$, it then follows the path $\ScTurnCurvingPath$ taking it to a configuration $\ScTurnArcStartConfiguration = (x_{\ScTurnArcStartCfgSubscript}, y_{\ScTurnArcStartCfgSubscript}, \theta_{\ScTurnArcStartCfgSubscript}, \maxcurvature)$.
The path $\ScTurnCurvingPath$ has initial and final curvatures $0$ and $\maxcurvature$, respectively.
The values $x_{\ScTurnArcStartCfgSubscript}$, $y_{\ScTurnArcStartCfgSubscript}$, and $\theta_{\ScTurnArcStartCfgSubscript}$ are those that result from following the curvature profile of $\ScTurnCurvingPath$ with a starting vehicle state $\ScTurnStartConfiguration$.

Once the vehicle has a curvature $\maxcurvature$, it then follows a circular arc path $\ScTurnArcPath$ with radius $\maxcurvature^{-1}$.
The circular arc starts at $(x_{\ScTurnArcStartCfgSubscript}, y_{\ScTurnArcStartCfgSubscript})$ and has its center at a distance $\maxcurvature^{-1}$ perpendicular to the orientation $\theta_{\ScTurnArcStartCfgSubscript}$ at point $(x_{\ScTurnArcStartCfgSubscript}, y_{\ScTurnArcStartCfgSubscript})$.
Its center is given by
\begin{align} \label{eq:circle_center}
(x_\Omega, y_\Omega) = (x_{\ScTurnArcStartCfgSubscript} - \maxcurvature^{-1} \sin \theta_{\ScTurnArcStartCfgSubscript}, y_{\ScTurnArcStartCfgSubscript} + \maxcurvature^{-1} \cos \theta_{\ScTurnArcStartCfgSubscript}).
\end{align}

The last path segment $\ScTurnDecurvingPath$ departs from the circular arc and it brings the vehicle to a configuration $\ScTurnEndConfiguration$.
Configuration $\ScTurnEndConfiguration$ depends on the point of departure from the circular arc, $\ScTurnArcEndConfiguration$.
However, it always lies in a circle $\Omega$, which has the same center as the circular arc $(x_\Omega, y_\Omega)$ in~\eqref{eq:circle_center}.

\def\AuxiliaryCircle{\Omega'}

In order to find the radius of circle $\Omega$, we first assume an auxiliary circular arc to be centered at ${(x_{\AuxiliaryCircle} , y_{\AuxiliaryCircle}) = (0, \maxcurvature^{-1})}$.
We assume a departure configuration from the circle at $(0, 0, 0, \maxcurvature)$.
Then, by following the path given by a curvature profile with initial and final curvatures $\maxcurvature$ and $\kappa_{\ScTurnEndCfgSubscript}$, we will end at a configuration $\textbf{q}_{\ScTurnEndCfgSubscript} = (x_{\ScTurnEndCfgSubscript}, y_{\ScTurnEndCfgSubscript}, \theta_{\ScTurnEndCfgSubscript}, \kappa_{\ScTurnEndCfgSubscript})$.
$\textbf{q}_{\ScTurnEndCfgSubscript}$ is a configuration located at an auxiliary $\AuxiliaryCircle$ circle (the auxiliary equivalent of the $\Omega$ circle), that has the same center as the circular arc.
Thus we compute the radius of $\AuxiliaryCircle$, which is equal to the radius of $\Omega$, as
\begin{align*}
r = \sqrt{ ( x_{\ScTurnEndCfgSubscript} - x_{\AuxiliaryCircle} )^2 + ( y_{\ScTurnEndCfgSubscript} - y_{\AuxiliaryCircle} )^2 } = \sqrt{ x_{\ScTurnEndCfgSubscript}^2 + ( y_{\ScTurnEndCfgSubscript} - \maxcurvature^{-1} )^2 }.
\end{align*}

An additional angle $\mu$ is defined as the difference between $\theta_{\ScTurnEndCfgSubscript}$ and the tangential angle to $\Omega$ at configuration $\ScTurnEndConfiguration$. It is computed using the previous auxiliary circular arc as
\begin{align}
\mu = \arctan \left( \frac{y_{\ScTurnEndCfgSubscript} - \maxcurvature^{-1} }{x_{\ScTurnEndCfgSubscript}} \right) + \frac{\pi}{2} - \theta_{\ScTurnEndCfgSubscript}.
\end{align}

Thus, given a certain initial configuration $\ScTurnStartConfiguration$, the possible positions of the ending configuration $\ScTurnEndConfiguration$, resulting from a combination of a cubic curvature path, a circular arc, and another cubic curvature path, \textit{i.e.}, an SC turn, lie on a circle $\Omega$.
The possible $\theta_{\ScTurnEndCfgSubscript}$ orientations of these configurations are given by the tangential angle at the circle plus $\mu$.

\subsection{Connecting Sharpness Continuous Turns}
\label{subsec:connecting_sharpness_continuous_turns}

An SC path between start and goal configurations $\startconfiguration$ and $\goalconfiguration$ can be found by connecting two SC turns.
An SC path consists of three elements:
\begin{itemize}
\item an SC turn starting at the start configuration $\startconfiguration$ and ending at a configuration $\TangentExitConfiguration$ with null curvature,
\item a line segment starting at $\TangentExitConfiguration$ and ending at $\TangentEntryConfiguration$,
\item an SC turn starting at configuration $\TangentEntryConfiguration$ with null curvature, and ending at the goal configuration $\goalconfiguration$.
\end{itemize}
Figure \ref{fig:sharpness_continuous_path} shows an example of an SC path, with the three elements described above.

\def\tikzpicturesscale{0.35}
\begin{figure}
  \centering
%     \resizebox {0.8\columnwidth} {!} {
%      \input{tikz/sc_path.tikz}
%      }
      \begin{tikzpicture}[scale=\tikzpicturesscale]

\def\radiusinnercircle{3}
\def\configurationlength{2}

\def\pathlinewidth{1.5}
\def\configurationslinewidth{1.5}

\def\startx{0}
\def\starty{0}
\def\startangle{0}

\def\endx{12}
\def\endy{5}
\def\endangle{-90}

\def\tangentanglestartcircle{-35}
\def\tangentangleendcircle{145}

%\draw [line width=\bracelinewidth, decorate,decoration={brace,mirror,amplitude=10pt}] ($(0,0) + (relativeoffset)$) -- ($(wp1) + (relativeoffset)$) node [midway,xshift=0.1in,yshift=-0.2in] {\tikzfontsscale $e_{1}$};

\coordinate (startnode) at (\startx, \starty);
\coordinate (endnode) at (\endx, \endy);

%\draw [line width=\primitiveinewidth, \primitivescolor, \primitivelinetype] (#2) to [out=#1 , in=-180+#1] ($(#2)+( {2*sqrt(2)*cos(#1)} , {2*sqrt(2)*sin(#1)} )$);

% Start configuration
\draw [ draw=black ,solid,line width=\configurationslinewidth,fill=black, preaction={-triangle 90,thin,draw,shorten >=-1mm} ] (startnode) to ($(startnode)+( {\configurationlength*cos(\startangle)} , {\configurationlength*sin(\startangle)} )$);
\node[below] (a) at (startnode) {\tikzfontsscale $\startconfiguration$};

% Left circle of start configuration
\coordinate (startcircleleft) at ($(startnode)+( {\radiusinnercircle*cos(\startangle+90)} , {\radiusinnercircle*sin(\startangle+90)} )$);
\draw [\AuxCircleDashPattern] (startcircleleft) circle (\radiusinnercircle);

% Right circle of start configuration
\coordinate (startcircleright) at ($(startnode)+( {\radiusinnercircle*cos(\startangle-90)} , {\radiusinnercircle*sin(\startangle-90)} )$);
\draw [\AuxCircleDashPattern] (startcircleright) circle (\radiusinnercircle);

% End configuration
\draw [ draw=black ,solid,line width=\configurationslinewidth,fill=black, preaction={-triangle 90,thin,draw,shorten >=-1mm} ] (endnode) to ($(endnode)+( {\configurationlength*cos(\endangle)} , {\configurationlength*sin(\endangle)} )$);
\node[right] (a) at (endnode) {\tikzfontsscale $\goalconfiguration$};

% Left circle of end configuration
\coordinate (endcircleleft) at ($(endnode)+( {\radiusinnercircle*cos(\endangle+90)} , {\radiusinnercircle*sin(\endangle+90)} )$);
\draw [\AuxCircleDashPattern] (endcircleleft) circle (\radiusinnercircle);

% Right circle of end configuration
\coordinate (endcircleright) at ($(endnode)+( {\radiusinnercircle*cos(\endangle-90)} , {\radiusinnercircle*sin(\endangle-90)} )$);
\draw [\AuxCircleDashPattern] (endcircleright) circle (\radiusinnercircle);

% Here we are computing the departure points from the circles (they will be connected over a straight line)
\coordinate (startcirclelefttangentpoint) at ($(startcircleleft)+( {\radiusinnercircle*cos(\tangentanglestartcircle)} , {\radiusinnercircle*sin(\tangentanglestartcircle)} )$);
\node[right] (a) at (startcirclelefttangentpoint) {\tikzfontsscale $\TangentExitConfiguration$};

\coordinate (endcirclerighttangentpoint) at ($(endcircleright)+( {\radiusinnercircle*cos(\tangentangleendcircle)} , {\radiusinnercircle*sin(\tangentangleendcircle)} )$);
\node[left] (a) at (endcirclerighttangentpoint) {\tikzfontsscale $\TangentEntryConfiguration$};

\draw [line width=\pathlinewidth] (startcirclelefttangentpoint) to (endcirclerighttangentpoint);

%\draw (0,0) arc (-30:30:2) ;
%Draws an arc that starts at (0,0), and would be the part of the circle from -30 degrees to +30 degrees of radius 2.

\draw [line width=\pathlinewidth] (startnode) arc (-90:\tangentanglestartcircle:\radiusinnercircle);

\draw [line width=\pathlinewidth] (endnode) arc (0:\tangentangleendcircle:\radiusinnercircle);

%\coordinate (circle1center) at ( 0 , 0 );
%\draw [densely dotted] (circle1center) circle (\radiusinnercircle);
%\coordinate (circle2center) at ( 0 , -10 );
%\draw [densely dotted] (circle2center) circle (\radiusinnercircle);

%\coordinate (circle3center) at ( 5 , 10 );
%\draw [densely dotted] (circle3center) circle (\radiusinnercircle);
%\coordinate (circle4center) at ( 15 , 10 );
%\draw [densely dotted] (circle4center) circle (\radiusinnercircle);

\end{tikzpicture}
  \caption{Sharpness continuous path example. The path consists of an SC turn between $\startconfiguration$ and $\TangentExitConfiguration$, a line segment from $\TangentExitConfiguration$ to $\TangentEntryConfiguration$, and an SC turn between $\TangentEntryConfiguration$ and $\goalconfiguration$. 
  The dashed circles correspond to the SC turns that can span from $\TangentExitConfiguration$ and $\TangentEntryConfiguration$.\label{fig:sharpness_continuous_path}}
\end{figure}
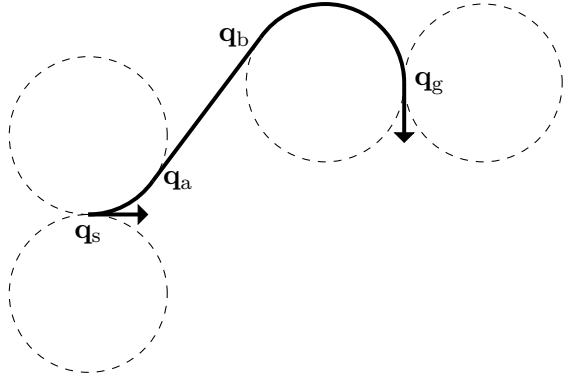

In order to connect two SC turns, we need to find the configurations $\TangentExitConfiguration$ and $\TangentEntryConfiguration$ that belong to the starting and ending SC turn possible departure configurations, and that can be connected with a line segment.
That is, $\TangentExitConfiguration$ and $\TangentEntryConfiguration$ must have the same orientation, \textit{i.e.}, $\theta_{\TangentExitCfgSubscript} = \theta_{\TangentEntryCfgSubscript}$.
Furthermore both must lie on a line segment with an inclination angle $\theta_{\TangentExitCfgSubscript}$.

As seen before, the possible set of departure configurations of an SC turn are located in a circle, and its orientations differ from the circle tangent by $\mu$.
Thus, to connect two SC turns, we need a way to connect two circles $\StartCircleOriginal$ and $\GoalCircleOriginal$ with arbitrary centers, radii, and $\mu$ values.

\def\tikzpicturesscale{0.45}

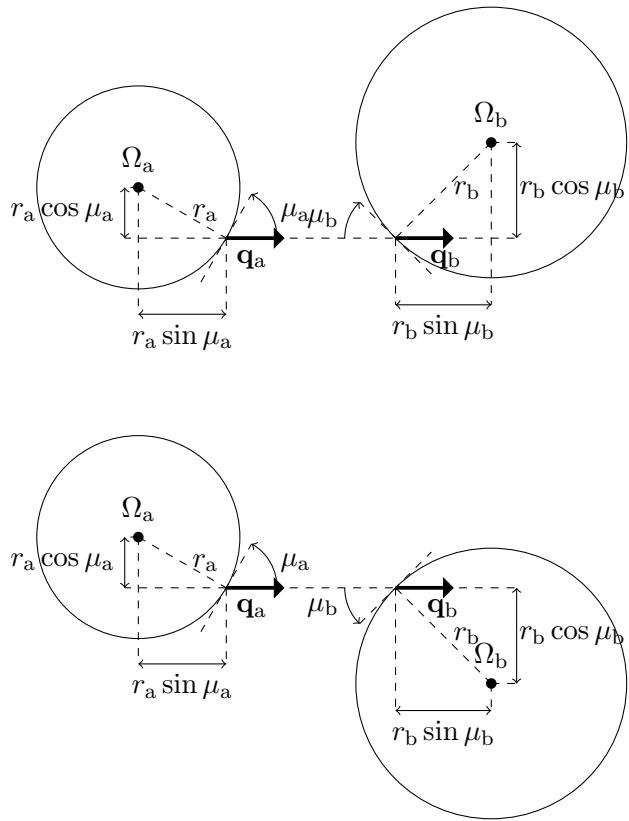
\begin{figure}
    \centering
    \begin{subfigure}[b]{0.5\textwidth}
      %  \resizebox {\columnwidth} {!} {
      %\input{tikz/external_tangents.tikz}
      %}
      \begin{tikzpicture}[scale=\tikzpicturesscale]

\def\radiusone{3}
\def\xcenterone{2}
\def\ycenterone{2}
\def\oangleone{-30}

\def\xqone{ { \xcenterone + \radiusone*cos( \oangleone ) } }
\def\yqone{ { \ycenterone + \radiusone*sin( \oangleone ) } }

\def\radiustwo{4}
\def\oangletwo{45}

\def\xqtwo{ \xqone + 5 }
\def\yqtwo{ \yqone }

\def\length_1{3}

%% First circle %% 

\coordinate (circle_1_center) at (\xcenterone ,\ycenterone );
\coordinate (q1) at (\xqone ,\yqone );

%\node at (circle_1_center) {\textbullet}; // Not really centered
\node[circle,fill=black,inner sep=0pt,minimum size=4pt,label=above:{\tikzfontsscale $\StartCircleAux$}] (a) at (circle_1_center) {};

\draw (circle_1_center) circle (\radiusone);

\draw[dashed] (circle_1_center) -- node [right] {\tikzfontsscale $r_{\StartCircleAuxSubscript}$ } (q1);

\draw[dashed] ($(q1) + ( { - 0.5*\length_1*cos( \oangleone + 90 ) } , { - 0.5*\length_1*sin( \oangleone + 90 ) } )$) -- ($(q1) +  ( { 0.5*\length_1*cos( \oangleone + 90 ) } , { 0.5*\length_1*sin( \oangleone + 90 ) } ) $);

% r1 cos mu1
\coordinate (r1cos1up) at ( \xqone - \radiusone , \ycenterone );
\coordinate (r1cos1down) at ($(q1) + ( -\radiusone , 0)$);
\draw[dashed] (circle_1_center) -- (r1cos1up);
\draw[dashed] (q1) -- (r1cos1down);
\draw[<->] (r1cos1up) -- node[left] {\tikzfontsscale $r_{\StartCircleAuxSubscript} \cos \mu_{\StartCircleAuxSubscript}$ } (r1cos1down);

% r1 sin mu1
%\draw[dashed] (circle_1_center) -- (\xcenterone , \ycenterone - 1.25*\radiusone );
\coordinate (r1sin1left) at ($ (circle_1_center) + (0, - 1.25*\radiusone) $);
\coordinate (r1sin1right) at ( \xqone , \ycenterone - 1.25*\radiusone );
\draw[dashed] (circle_1_center) -- (r1sin1left);
\draw[dashed] (q1) -- (r1sin1right);
\draw[<->] (r1sin1left) -- node[below] {\tikzfontsscale $r_{\StartCircleAuxSubscript} \sin \mu_{\StartCircleAuxSubscript}$ } (r1sin1right);

% q1
\path[draw=black,solid,line width=0.5mm,fill=black,
preaction={-triangle 90,thin,draw,shorten >=-1mm}
] (q1) -- node[below] {\tikzfontsscale $\mathbf{q}_{\StartCircleAuxSubscript}$ } ($(q1) + (1.5, 0)$);

% Arc
\draw[->] ($(q1) + (1.5, 0)$) arc (0:{90 - abs(\oangleone) }:1.5) node[midway, right]{\tikzfontsscale $\mu_{\StartCircleAuxSubscript}$};

%% Second circle %%
\coordinate (circletwocenter) at ({\xqtwo + \radiustwo*cos( \oangletwo )}, {\yqtwo + \radiustwo*sin( \oangletwo )});
%\coordinate (circle_2_center) at (\xcentertwo , \ycentertwo);
\coordinate (q2) at (\xqtwo ,\yqtwo );

% q2
\path[draw=black,solid,line width=0.5mm,fill=black,
preaction={-triangle 90,thin,draw,shorten >=-1mm}
] (q2) -- node[below] {\hspace{1.4em} \tikzfontsscale $\mathbf{q}_{\GoalCircleAuxSubscript}$ } ($(q2) + (1.5, 0)$);

\draw (circletwocenter) circle (\radiustwo);

\draw[dashed] (circletwocenter) -- node [right] {\tikzfontsscale $r_{\GoalCircleAuxSubscript}$ } (q2);

\node[circle,fill=black,inner sep=0pt,minimum size=4pt,label=above:{\tikzfontsscale $\GoalCircleAux$}] (a) at (circletwocenter) {};

\draw[dashed] ($(q2) + ( { - 0.5*\length_1*cos( \oangletwo + 90 ) } , { - 0.5*\length_1*sin( \oangletwo + 90 ) } )$) -- ($(q2) +  ( { 0.5*\length_1*cos( \oangletwo + 90 ) } , { 0.5*\length_1*sin( \oangletwo + 90 ) } ) $);

% r2 cos mu2
\coordinate (rtwocostwoup) at ($(circletwocenter) + (0.25*\radiusone, 0) $);
\coordinate (rtwocostwodown) at ($(q2) + ( { \radiustwo*cos( \oangletwo ) + 0.25*\radiusone } , 0) $);
\draw[dashed] (circletwocenter) -- (rtwocostwoup);
\draw[dashed] (q2) -- (rtwocostwodown);
\draw[<->] (rtwocostwoup) -- node[right] { \hspace{-0.25em}\tikzfontsscale $r_{\GoalCircleAuxSubscript} \cos \mu_{\GoalCircleAuxSubscript}$ } (rtwocostwodown);

% r2 sin mu2
\def\rtwosintworightx{ \xqtwo + \radiustwo*cos( \oangletwo ) }
\def\rtwosintworighty{ \yqtwo + \radiustwo*sin( \oangletwo ) - 1.2*\radiustwo }

\def\rtwosintwoleftx{ \xqtwo }
\def\rtwosintwolefty{ \yqtwo + \radiustwo*sin( \oangletwo ) - 1.2*\radiustwo }

\coordinate (rtwosintworight) at ($( { \rtwosintworightx }, { \rtwosintworighty } )$);
\coordinate (rtwosintwoleft) at ($( { \rtwosintwoleftx }, { \rtwosintwolefty } )$);
\draw[dashed] (circletwocenter) -- (rtwosintworight);
\draw[dashed] (q2) -- (rtwosintwoleft);
\draw[<->] (rtwosintwoleft) -- node[below] { \tikzfontsscale $r_{\GoalCircleAuxSubscript} \sin \mu_{\GoalCircleAuxSubscript}$ } (rtwosintworight);

% Arc
\draw[->] ($(q2) +(-1.5, 0)$) arc (180:{180 - abs(\oangletwo) }:1.5) node[midway, left]{\tikzfontsscale $\mu_{\GoalCircleAuxSubscript}$};

\draw[dashed] (q1) -- (q2);

\end{tikzpicture}
    \end{subfigure}
    \\ % This is here so that the vspace appears between the two figures
    \vspace{0.5cm}
    \begin{subfigure}[b]{0.5\textwidth}
      %  \resizebox {\columnwidth} {!} {
      %\input{tikz/internal_tangents.tikz}
      %}
      \begin{tikzpicture}[scale=\tikzpicturesscale]

\def\radiusone{3}
\def\xcenterone{2}
\def\ycenterone{2}
\def\oangleone{-30}

\def\xqone{ { \xcenterone + \radiusone*cos( \oangleone ) } }
\def\yqone{ { \ycenterone + \radiusone*sin( \oangleone ) } }

\def\radiustwo{4}
\def\oangletwo{45}

\def\xqtwo{ \xqone + 5 }
\def\yqtwo{ \yqone }

\def\length_1{3}

%% First circle %% 

\coordinate (circle_1_center) at (\xcenterone ,\ycenterone );
\coordinate (q1) at (\xqone ,\yqone );

%\node at (circle_1_center) {\textbullet}; // Not really centered
\node[circle,fill=black,inner sep=0pt,minimum size=4pt,label=above:{\tikzfontsscale $\StartCircleAux$}] (a) at (circle_1_center) {};

\draw (circle_1_center) circle (\radiusone);

\draw[dashed] (circle_1_center) -- node [right] {\tikzfontsscale  $r_{\StartCircleAuxSubscript}$ } (q1);

\draw[dashed] ($(q1) + ( { - 0.5*\length_1*cos( \oangleone + 90 ) } , { - 0.5*\length_1*sin( \oangleone + 90 ) } )$) -- ($(q1) +  ( { 0.5*\length_1*cos( \oangleone + 90 ) } , { 0.5*\length_1*sin( \oangleone + 90 ) } ) $);

% r1 cos mu1
\coordinate (r1cos1up) at ( \xqone - \radiusone , \ycenterone );
\coordinate (r1cos1down) at ($(q1) + ( -\radiusone , 0)$);
\draw[dashed] (circle_1_center) -- (r1cos1up);
\draw[dashed] (q1) -- (r1cos1down);
\draw[<->] (r1cos1up) -- node[left] {\tikzfontsscale  $r_{\StartCircleAuxSubscript} \cos \mu_{\StartCircleAuxSubscript}$ } (r1cos1down);

% r1 sin mu1
%\draw[dashed] (circle_1_center) -- (\xcenterone , \ycenterone - 1.25*\radiusone );
\coordinate (r1sin1left) at ($ (circle_1_center) + (0, - 1.25*\radiusone) $);
\coordinate (r1sin1right) at ( \xqone , \ycenterone - 1.25*\radiusone );
\draw[dashed] (circle_1_center) -- (r1sin1left);
\draw[dashed] (q1) -- (r1sin1right);
\draw[<->] (r1sin1left) -- node[below] {\tikzfontsscale $r_{\StartCircleAuxSubscript} \sin \mu_{\StartCircleAuxSubscript}$ } (r1sin1right);

% q1
\path[draw=black,solid,line width=0.5mm,fill=black,
preaction={-triangle 90,thin,draw,shorten >=-1mm}
] (q1) -- node[below] {\tikzfontsscale $\mathbf{q}_{\StartCircleAuxSubscript}$ } ($(q1) + (1.5, 0)$);

% Arc
\draw[->] ($(q1) +(1.5, 0)$) arc (0:{90 - abs(\oangleone) }:1.5) node[midway, right]{\tikzfontsscale $\mu_{\StartCircleAuxSubscript}$};

%% Second circle %%
\coordinate (circletwocenter) at ({\xqtwo + \radiustwo*cos( \oangletwo )}, {\yqtwo - \radiustwo*sin( \oangletwo )});
%\coordinate (circle_2_center) at (\xcentertwo , \ycentertwo);
\coordinate (q2) at (\xqtwo ,\yqtwo );

% q2
\path[draw=black,solid,line width=0.5mm,fill=black,
preaction={-triangle 90,thin,draw,shorten >=-1mm}
] (q2) -- node[below] {\hspace{1.2em} \tikzfontsscale $\mathbf{q}_{\GoalCircleAuxSubscript}$ } ($(q2) + (1.5, 0)$);

\draw (circletwocenter) circle (\radiustwo);

\draw[dashed] (circletwocenter) -- node [right] {\tikzfontsscale $r_{\GoalCircleAuxSubscript}$ } (q2);

\node[circle,fill=black,inner sep=0pt,minimum size=4pt,label=above:{\tikzfontsscale $\GoalCircleAux$}] (a) at (circletwocenter) {};

\draw[dashed] ($(q2) + ( { - 0.5*\length_1*cos( \oangletwo ) } , { - 0.5*\length_1*sin( \oangletwo ) } )$) -- ($(q2) +  ( { 0.5*\length_1*cos( \oangletwo ) } , { 0.5*\length_1*sin( \oangletwo ) } ) $);

% r2 cos mu2
\coordinate (rtwocostwoup) at ($(circletwocenter) + (0.25*\radiusone, 0) $);
\coordinate (rtwocostwodown) at ($(q2) + ( { \radiustwo*cos( \oangletwo ) + 0.25*\radiusone } , 0) $);
\draw[dashed] (circletwocenter) -- (rtwocostwoup);
\draw[dashed] (q2) -- (rtwocostwodown);
\draw[<->] (rtwocostwoup) -- node[right] {\hspace{-0.6em} \tikzfontsscale $r_{\GoalCircleAuxSubscript} \cos \mu_{\GoalCircleAuxSubscript}$ } (rtwocostwodown);

% r2 sin mu2
\def\rtwosintworightx{ \xqtwo + \radiustwo*cos( \oangletwo ) }
\def\rtwosintworighty{ \yqtwo - \radiustwo*sin( \oangletwo ) - 0.2*\radiustwo }

\def\rtwosintwoleftx{ \xqtwo }
\def\rtwosintwolefty{ \yqtwo - \radiustwo*sin( \oangletwo ) - 0.2*\radiustwo }

\coordinate (rtwosintworight) at ($( { \rtwosintworightx }, { \rtwosintworighty } )$);
\coordinate (rtwosintwoleft) at ($( { \rtwosintwoleftx }, { \rtwosintwolefty } )$);
\draw[dashed] (circletwocenter) -- (rtwosintworight);
\draw[dashed] (q2) -- (rtwosintwoleft);
\draw[<->] (rtwosintwoleft) -- node[below] { \tikzfontsscale $r_{\GoalCircleAuxSubscript} \sin \mu_{\GoalCircleAuxSubscript}$ } (rtwosintworight);

% Arc
\draw[->] ($(q2) + (-1.5, 0)$) arc (-180:{-180 + abs(\oangletwo) }:1.5) node[midway, left]{\tikzfontsscale $\mu_{\GoalCircleAuxSubscript}$};

\draw[dashed] (q1) -- (q2);

\end{tikzpicture}
      \end{subfigure}
    \caption{Computing the external (top) and internal (bottom) tangents between two circles.\vspace*{-5mm}}\label{fig:tangents}
\end{figure}

We first assume two auxiliary circles $\StartCircleAux$ and $\GoalCircleAux$, as depicted in Figure \ref{fig:tangents} (top).
$\StartCircleAux$ and $\GoalCircleAux$ have the same radii and $\mu$ values as the original circles $\StartCircleOriginal$ and $\GoalCircleOriginal$.
$\StartCircleAux$ is centered at $(0, 0)$ and $\GoalCircleAux$ is located so that $\TangentExitConfiguration$ and $\TangentEntryConfiguration$ are collinear.
We are interested in finding the center of $\GoalCircleAux = (x_{\GoalCircleAux} , y_{\GoalCircleAux} )$.
From Figure \ref{fig:tangents} (top) it can be seen that
\begin{align}
y_{\GoalCircleAuxSubscript} = -r_{\StartCircleAuxSubscript} \cos \mu_{\StartCircleAuxSubscript} + r_{\GoalCircleAuxSubscript} \cos \mu_{\GoalCircleAuxSubscript}.
\end{align}

We assume that the distance $r(\StartCircleAux, \GoalCircleAux)$ between the circle centers is the same as the distance between the original circles $r(\StartCircleOriginal, \GoalCircleOriginal)$. We then have
\begin{align}
x_{\GoalCircleAux} = \sqrt{ r(\StartCircleOriginal, \GoalCircleOriginal)^2 - y_{\GoalCircleAux}^2 }.
\end{align}

\sloppy We know that ${\TangentExitConfiguration = (r_{\StartCircleAuxSubscript} \sin \mu_{\StartCircleAuxSubscript}, -r_{\StartCircleAuxSubscript} \cos \mu_{\StartCircleAuxSubscript}, 0, 0)}$, and that ${\TangentEntryConfiguration = ( x_{\GoalCircleAux} - r_{\GoalCircleAuxSubscript} \sin \mu_{\GoalCircleAuxSubscript} , y_{\GoalCircleAux} - r_{\GoalCircleAuxSubscript} \cos \mu_{\GoalCircleAuxSubscript} , 0, 0)}$.
To find these configurations in the original circles $\StartCircleOriginal$ and $\GoalCircleOriginal$, we need to first apply a rotation ${\Delta_\theta = \arctan ( y_{\GoalCircleOriginal} - y_{\StartCircleOriginal} , x_{\GoalCircleOriginal} - x_{\StartCircleOriginal} )}$ to $\TangentExitConfiguration$ and $\TangentEntryConfiguration$.
We then translate these configurations by $(\Delta_x, \Delta_y) = ( x_{\StartCircleOriginal}, y_{\StartCircleOriginal} )$.
The resulting rotated and translated configurations correspond to the desired tangent configurations between the circles $\StartCircleOriginal$ and $\GoalCircleOriginal$.

The above procedure finds the departure configurations between two counter clockwise (left steering) SC turns, shown in Figure~\ref{fig:tangents} (top).
An analogous procedure can be used to find the possible departure configurations between any combination of clockwise (right steering) and counter clockwise turns, as shown in Figure~\ref{fig:tangents} (bottom). 
This procedures are valid if the found tangent configurations $\TangentExitConfiguration$ and $\TangentEntryConfiguration$ do not lie inside the circles $\GoalCircleAux$ and $\StartCircleAux$, respectively. 

\subsection{Finding the Shortest SC Path}
\label{subsec:sc_shortest_path}

In order to find the shortest SC path between two configurations $\startconfiguration$ and $\goalconfiguration$, we need to compute all the possible SC turns that can be spanned from these configurations.
The SC turns are then connected, in order to generate possible SC paths.
The process is detailed below.

Each of the configurations $\startconfiguration$ and $\goalconfiguration$ can span a total of four SC turns, depending if the vehicle is moving forwards or backwards, or if it is turning left or right.
The possible SC turns that span from $\startconfiguration$ and $\goalconfiguration$ are shown in Figure~\ref{fig:sharpness_continuous_path} as dashed circles (forward and backward SC turns are equivalent, so they lie on top of each other).
Figure~\ref{fig:sharpness_continuous_turn} shows an SC turn which assumes a vehicle moving forward and turning left.
The method explained in section \ref{subsec:sharpness_continuous_turns} can be readily used to obtain SC turns moving forward, independent of the direction they are turning.
The procedure to obtain an SC turn moving backwards is analogous.

There are a total of 16 possible SC paths between the two sets of 4 SC turns spanned from $\startconfiguration$ and $\goalconfiguration$.
Each path is found by computing the SC path, resulting from connecting two SC turns, as detailed in~\ref{subsec:connecting_sharpness_continuous_turns}.
Each SC path length is evaluated, and the shortest is selected as the solution.

%%%%%%%%%%%%%%%%%%%%%%%%%%%%%%%%
\section{Cubic Curvature Paths}
\label{sec:cubic_curvature_paths}

Cubic curvature paths are a building block of an SC turn.
They are used as transition paths connecting configurations to and from an arc circle (paths $\ScTurnCurvingPath$ and $\ScTurnDecurvingPath$ in~\ref{subsec:sharpness_continuous_turns}), and they guarantee the sharpness continuity of the whole path.

\subsection{Introduction}

\def\PathLength{s_{\mathrm{f}}}
\def\InitialCurvature{\kappa_{\mathrm{i}}}
\def\InitialSharpness{\alpha_{\mathrm{i}}}
\def\FinalCurvature{\kappa_{\mathrm{f}}}
\def\FinalSharpness{\alpha_{\mathrm{f}}}

Cubic curvature paths are defined as paths with a cubic curvature profile $\kappa(s) = a_3 s^3 + a_2 s^2 + a_1 s + a_0$, where $s$ is the length along the path.
A cubic curvature profile is the minimum degree polynomial that allows to define arbitrary initial and final curvatures, $\InitialCurvature$ and $\FinalCurvature$, and sharpnesses $\InitialSharpness$ and $\FinalSharpness$. The sharpness profile of these paths is given by:
\begin{align} 
\label{eq:sharpness_equation}
\alpha(s) = \frac{\partial \kappa(s) }{ \partial s } = 3 a_3 s^2 + 2 a_2 s + a_1.
\end{align} 
In order to find the parameters of the cubic polynomial, we use the initial and final constraints:
\begin{equation}
\label{eq:cubic_curvature_linear_system}
\kappa(0) = \InitialCurvature,
\quad
\alpha(0) = \InitialSharpness,
\quad
\kappa(\PathLength) = \FinalCurvature,
\quad
\alpha(\PathLength) = \FinalSharpness,
\end{equation}
where $\PathLength$ is the path length, and is itself an unknown.

We want to ensure sharpness continuity, so we need to set the initial and final sharpness values, $\InitialSharpness$ and $\FinalSharpness$, to zero.
This allows us to stitch together cubic curvature paths with line and arc segments, which have null sharpness, while ensuring sharpness continuity.
It should be noted that \eqref{eq:cubic_curvature_linear_system} does not take into account steering limitations, thus, infeasible paths can be generated.
In the following we address this issue.

\subsection{Ensuring Steering Rate and Acceleration Constraints}
\label{subsec:ensuring_rate_and_accel_constraints}

In order to have a feasible path, we need to ensure that a vehicle can follow it while complying with its steering constraints.
If we make both $\InitialCurvature$ and $\FinalCurvature$ smaller in magnitude than $\maxcurvature$, we ensure that the path always has a steering angle magnitude smaller than $\maxsteering$.

The steering angle profile corresponding to the cubic curvature path is first computed from the path curvature using \eqref{eq:curvature_from_steering}.
Assuming then that the vehicle is following the path at a given fixed velocity $\mathbf{v}$, the steering angle rate and acceleration profiles are computed.
Both profiles have a peak magnitude rate $\peaksteeringdot$ and acceleration $\peaksteeringdotdot$.
In case $\peaksteeringdot$ is larger than the allowed maximum steering rate $\maxsteeringdot$ the length $\PathLength$ needs to be increased so that $\peaksteeringdot = \maxsteeringdot$.
This can be achieved by simply scaling the path length by $\peaksteeringdot/\maxsteeringdot$.
Similarly if $\peaksteeringdotdot > \maxsteeringdotdot$, we scale the path length by a scaling factor of $\surd ( \peaksteeringdotdot / \maxsteeringdotdot )$.
To guarantee that the path respects both steering rate and acceleration limitations, we need to scale its length by the greater of the scaling factors.
Once the new path length $\PathLength$ is computed, the cubic curvature path is recomputed, by solving~\eqref{eq:cubic_curvature_linear_system}.

%%%%%%%%%%%%%%%%%
\section{Results}
\label{sec:results}

\subsection{Convergence of SC Paths to Dubins Paths}

As previously mentioned, Dubins paths are proven to be optimal in terms of length.
SC paths are, however, longer than the Dubins path.
This happens because SC turns have longer turning radii than a circular arc with radius $\maxcurvature^{-1}$.

\begin{figure}
  \centering
      \resizebox {\columnwidth} {!} {
      	\input{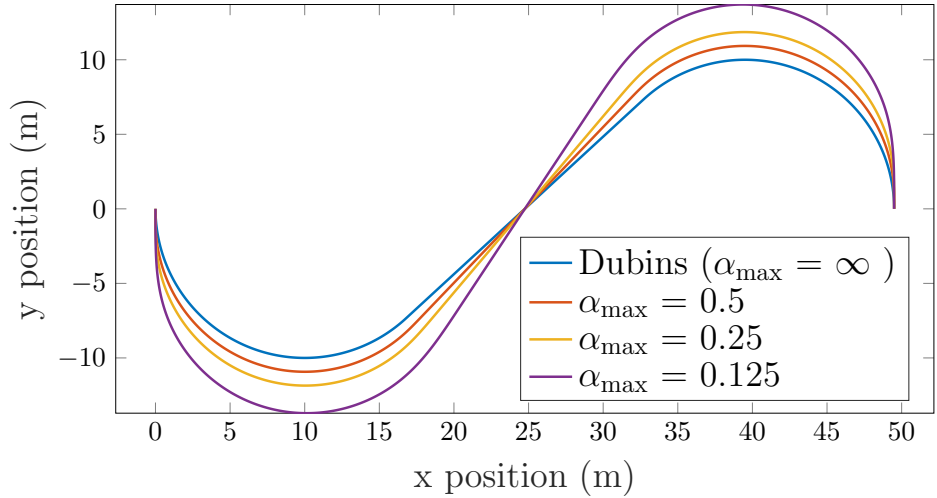}
      }
  \caption{SC paths with increasing sharpness $\alpha$ converge to the length optimal Dubins path. \label{fig:sharpness_increasing}\vspace*{-3mm}}
\end{figure}

Figure~\ref{fig:sharpness_increasing} shows the Dubins path for a given start and goal configurations $\startconfiguration$ and $\goalconfiguration$.
Overlayed are SC paths with different maximum sharpness $\alpha_{\max}$ for the same configurations $\startconfiguration$ and $\goalconfiguration$.
It is seen that the greater the maximum sharpness $\alpha_{\max}$ of the SC paths is, \textit{i.e.}, the greater the achievable steering rate and accelerations of the vehicle, the closer it approaches the Dubins path.
This is somewhat intuitive, as increasing $\alpha_{\max}$ results in increasing the rate of change of the curvature profile.
If $\alpha_{\max} \rightarrow \infty$, then the curvature changes would be immediate, and the SC path would be equivalent to the Dubins path, and as such, length optimal.

\subsection{Notes on Computational Cost}

As previously mentioned, given two configurations $\startconfiguration$ and $\goalconfiguration$, the SC method computes all possible 16 SC turns and how they can be connected.
The connection process, as detailed in section \ref{subsec:connecting_sharpness_continuous_turns}, is computationally cheap.
The bulk of processing comes from finding all 16 possible SC turns.

As seen in section \ref{subsec:sharpness_continuous_turns}, an SC turn depends on the cubic curvature paths that are part of it.
In order to evaluate these paths, one has to generate their curvature profiles from the given initial and final constraints.
To comply with the steering constraints, a numerical evaluation of a steering profile must be done, in order to find the path length scaling factors, as detailed in section \ref{subsec:ensuring_rate_and_accel_constraints}.
When one has the desired curvature profile, the orientations $\theta$ can be obtained analytically.
The $x$ and $y$ positions of the path are found by solving the vehicle model equations, using an Euler method, which has a high computational cost.
This cost can be greatly reduced using precomputations, as detailed in the following.

\subsection{Precomputation of Cubic Curvature Paths}

As previously stated, the SC method computation speed is limited by the generation of the cubic curvature paths.
Depending on the application, some assumptions can be made that greatly improve the computation speed, by allowing the precomputation of the cubic curvature paths to be used.

If one assumes that the start and end configurations $\startconfiguration$ and $\goalconfiguration$ always have null curvatures, then one can compute, in an initialization procedure, all the possible cubic curvature paths starting and ending at curvatures $\kappa = 0, \pm \maxcurvature$.
Thus we skip the expensive generation of cubic curvature paths needed to find out the possible SC turns.
In order to generate an SC turn, one just has to use the precomputed paths and apply rotations and translations on them.

The precomputation of paths can still be achieved, without limiting the start and goal configurations $\startconfiguration$ and $\goalconfiguration$ to have null curvature.
In fact, one can allow $\startconfiguration$ and $\goalconfiguration$ to have curvature values belonging to finite discrete set.

\subsection{Timing Evaluation}

We test the steering method, measuring its computational speed for several problem instances. The method is implemented in \texttt{C++} and running on a Linux Mint distribution. The computer used is equipped with an Intel Core i7-6820HQ Processor running at 2.70 GHz, and with 16,0 GB of RAM.

We generate 1000 random pairs of start and goal configuration queries.
Each query is repeated 100 times to get a better estimate of the average computational time.
The start and goal configurations of each query are generated by sampling the $x$ and $y$ coordinates from a uniform random distribution between $-50$ and $50$.
The orientations sampled from the interval $[-\pi, \pi]$, and the curvatures from a discrete equispaced set of 11 curvatures $[-\maxcurvature, \ldots, 0, \ldots, \maxcurvature]$.

When making use of precomputations, we get an average time for finding a solution path of $70 \si{\micro \second}$, while without precomputations we get an average time of $12 \si{\milli \second}$.
The precomputations greatly decrease the computational time.
These results indicate that the steering method is extremely inexpensive when using precomputations.
Even without precomputations, the method runs in few milliseconds, making it suitable for real-time applications.

\subsection{Simulations}

A simulation test is run in order to understand how the proposed paths affect the performance of a vehicle tracking them.
A kinematic vehicle model coupled with a detailed steering actuator model are used to simulate a vehicle.

In the test, two paths consisting of a straight segment, a turn, and a straight segment are generated.
The first path is a CC path~\cite{fraichard2004reeds}, while the second is our proposed SC path.
Both paths abide by the same maximum steering angle magnitude $\maxsteering$ and steering angle rate $\maxsteeringdot$ constraints.
Additionally, the SC path respects also the steering angle acceleration $\maxsteeringdotdot$ constraint, unlike the CC paths.

The simulation assumes a steering actuator that is limited in terms of achievable steering angle magnitude, rate, and acceleration.
The steering angle is controlled making use of a PID controller, which receives a steering angle reference, and actuates on the steering angle torque.
The PID controller was tuned to achieve a step response with a relatively fast settling time and little overshoot.
The steering angle reference is provided from a high-level path tracking controller.
The high-level controller consists of a feedforward part and a feedback part.
The feedforward part is obtained by finding the closest path point, and getting the corresponding steering angle reference at that point.
The feedback part is a proportional controller regulating both lateral and heading errors.
Such a controller is a simple implementation commonly used in path tracking applications. %\highlight{Reference?}

\begin{figure}
  \centering
      \resizebox {\columnwidth} {!} { \input{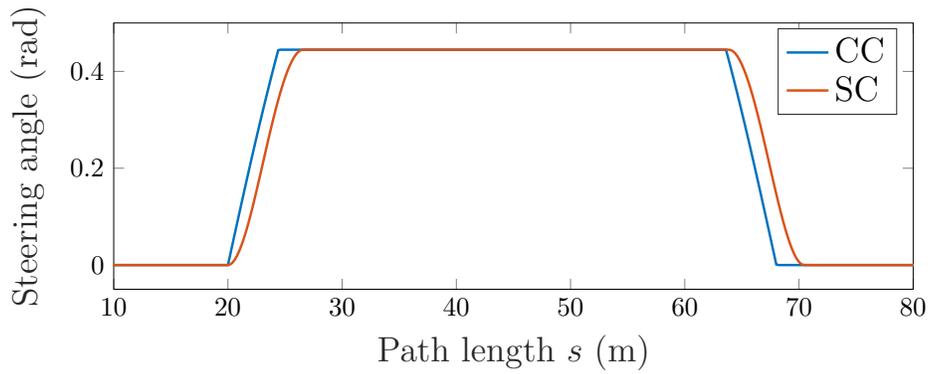} }
  \caption{Steering reference profiles used in simulation. \label{fig:sim_steering_reference}}
\end{figure}

Figure \ref{fig:sim_steering_reference} shows the steering reference profiles of the paths to be tracked.
The difference between them is in the shape of the increasing and decreasing sections of the steering angle.
In the CC case, the steering angle change follows a linear profile while in the SC it follows a cubic profile.

\begin{figure}
  \centering
      \resizebox {\columnwidth} {!} { \input{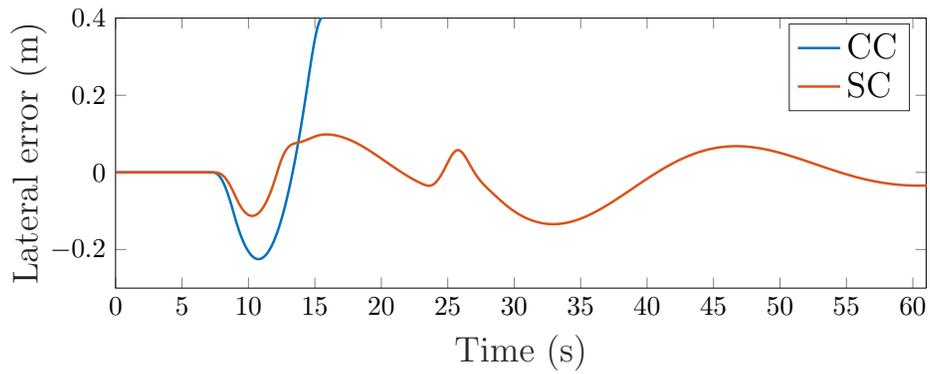} }
  \caption{Lateral error when tracking a CC and an SC path. When tracking the CC path, the controller becomes unstable, resulting in an error that grows indefinitely, and out of scope of the graph. The SC path tracking is seen to be stable.\label{fig:sim_errors}}
\end{figure}

Figure \ref{fig:sim_errors} shows the lateral and heading errors when the vehicle tracks both paths.
The vehicle is initially placed at the start of the path, and it follows the first straight segment perfectly.
However, when the turning section starts, a deviation from the path begins to arise.
The feedback part is then responsible for trying to regulate the errors to zero.
Shortly after the turn begins, the CC case becomes unstable.
On the other hand, the SC case is stable, and its error converges to zero.
The error profiles show that the controller performance is worse when tracking CC paths.

\begin{figure}
  \centering
      \resizebox {\columnwidth} {!} { % This file was created by matlab2tikz.
\definecolor{mycolor1}{rgb}{0.00000,0.44700,0.74100}%
\definecolor{mycolor2}{rgb}{0.85000,0.32500,0.09800}%
\begin{tikzpicture}

\begin{axis}[%
width=4.521in,
height=2.26in,
at={(0.758in,1.134in)},
scale only axis,
xmin=0,
xmax=40,
xlabel style={font=\color{white!15!black}},
xlabel={\tikzmatlabfontsscale Time (s)},
ymin=-0.45,
ymax=0.45,
ylabel style={font=\color{white!15!black}},
ylabel={\tikzmatlabfontsscale Lateral acceleration (m/s$^2$)},
axis background/.style={fill=white},
legend style={legend cell align=left, align=left, draw=white!15!black}
]
\addplot [color=mycolor1, line width=1.0pt]
  table[row sep=crcr]{%
0	0\\
0.1	0\\
0.2	0\\
0.3	0\\
0.4	0\\
0.5	0\\
0.6	0\\
0.7	0\\
0.8	0\\
0.9	0\\
1	0\\
1.1	0\\
1.2	0\\
1.3	0\\
1.4	0\\
1.5	0\\
1.6	0\\
1.7	0\\
1.8	0\\
1.9	0\\
2	0\\
2.1	0\\
2.2	0\\
2.3	0\\
2.4	0\\
2.5	0\\
2.6	0\\
2.7	0\\
2.8	0\\
2.9	0\\
3	0\\
3.1	0\\
3.2	0\\
3.3	0\\
3.4	0\\
3.5	0\\
3.6	0\\
3.7	0\\
3.8	0\\
3.9	0\\
4	0\\
4.1	0\\
4.2	0\\
4.3	0\\
4.4	0\\
4.5	0\\
4.6	0\\
4.7	0\\
4.8	0\\
4.9	0\\
5	0\\
5.1	0\\
5.2	0\\
5.3	0\\
5.4	0\\
5.5	0\\
5.6	0\\
5.7	0\\
5.8	0\\
5.9	0\\
6	0\\
6.1	0\\
6.2	0\\
6.3	0\\
6.4	0\\
6.5	0\\
6.6	0\\
6.7	0\\
6.8	0\\
6.9	0\\
7	0\\
7.1	0\\
7.2	0\\
7.3	0.420328944901333\\
7.4	0.420112486119246\\
7.5	0.389647208314116\\
7.6	0.418850758206747\\
7.7	0.417786847045321\\
7.8	0.416442969925104\\
7.9	0.385222825221921\\
8	0.41300639795152\\
8.1	0.410862633663876\\
8.2	0.408461918400129\\
8.3	0.405910867017066\\
8.4	0.403028718500091\\
8.5	0.399914624458481\\
8.6	0.396701095156029\\
8.7	0.393159920784064\\
8.8	0.361665185476513\\
8.9	0\\
9	0\\
9.1	0\\
9.2	0\\
9.3	0\\
9.4	0\\
9.5	0\\
9.6	0\\
9.7	0\\
9.8	0\\
9.9	0\\
10	0\\
10.1	0\\
10.2	0\\
10.3	0\\
10.4	0\\
10.5	0\\
10.6	0\\
10.7	0\\
10.8	0\\
10.9	0\\
11	0\\
11.1	0\\
11.2	0\\
11.3	0\\
11.4	0\\
11.5	0\\
11.6	0\\
11.7	0\\
11.8	0\\
11.9	0\\
12	0\\
12.1	0\\
12.2	0\\
12.3	0\\
12.4	0\\
12.5	0\\
12.6	0\\
12.7	0\\
12.8	0\\
12.9	0\\
13	0\\
13.1	0\\
13.2	0\\
13.3	0\\
13.4	0\\
13.5	0\\
13.6	0\\
13.7	0\\
13.8	0\\
13.9	0\\
14	0\\
14.1	0\\
14.2	0\\
14.3	0\\
14.4	0\\
14.5	0\\
14.6	0\\
14.7	0\\
14.8	0\\
14.9	0\\
15	0\\
15.1	0\\
15.2	0\\
15.3	0\\
15.4	0\\
15.5	0\\
15.6	0\\
15.7	0\\
15.8	0\\
15.9	0\\
16	0\\
16.1	0\\
16.2	0\\
16.3	0\\
16.4	0\\
16.5	0\\
16.6	0\\
16.7	0\\
16.8	0\\
16.9	0\\
17	0\\
17.1	0\\
17.2	0\\
17.3	0\\
17.4	0\\
17.5	0\\
17.6	0\\
17.7	0\\
17.8	0\\
17.9	0\\
18	0\\
18.1	0\\
18.2	0\\
18.3	0\\
18.4	0\\
18.5	0\\
18.6	0\\
18.7	0\\
18.8	0\\
18.9	0\\
19	0\\
19.1	0\\
19.2	0\\
19.3	0\\
19.4	0\\
19.5	0\\
19.6	0\\
19.7	0\\
19.8	0\\
19.9	0\\
20	0\\
20.1	0\\
20.2	0\\
20.3	0\\
20.4	0\\
20.5	0\\
20.6	0\\
20.7	0\\
20.8	0\\
20.9	0\\
21	0\\
21.1	0\\
21.2	0\\
21.3	0\\
21.4	0\\
21.5	0\\
21.6	0\\
21.7	0\\
21.8	0\\
21.9	0\\
22	0\\
22.1	0\\
22.2	0\\
22.3	0\\
22.4	0\\
22.5	0\\
22.6	0\\
22.7	0\\
22.8	0\\
22.9	0\\
23	0\\
23.1	-0.388727911768988\\
23.2	-0.364409078545217\\
23.3	-0.3959598110539\\
23.4	-0.399334856997552\\
23.5	-0.402489370930868\\
23.6	-0.405313253573641\\
23.7	-0.408006763705625\\
23.8	-0.410452565095625\\
23.9	-0.412642751536233\\
24	-0.414505915994174\\
24.1	-0.416173863532646\\
24.2	-0.417567283465489\\
24.3	-0.418681480483155\\
24.4	-0.419487897305748\\
24.5	-0.42004350133628\\
24.6	-0.420311451959596\\
24.7	0\\
24.8	0\\
24.9	0\\
25	0\\
25.1	0\\
25.2	0\\
25.3	0\\
25.4	0\\
25.5	0\\
25.6	0\\
25.7	0\\
25.8	0\\
25.9	0\\
26	0\\
26.1	0\\
26.2	0\\
26.3	0\\
26.4	0\\
26.5	0\\
26.6	0\\
26.7	0\\
26.8	0\\
26.9	0\\
27	0\\
27.1	0\\
27.2	0\\
27.3	0\\
27.4	0\\
27.5	0\\
27.6	0\\
27.7	0\\
27.8	0\\
27.9	0\\
28	0\\
28.1	0\\
28.2	0\\
28.3	0\\
28.4	0\\
28.5	0\\
28.6	0\\
28.7	0\\
28.8	0\\
28.9	0\\
29	0\\
29.1	0\\
29.2	0\\
29.3	0\\
29.4	0\\
29.5	0\\
29.6	0\\
29.7	0\\
29.8	0\\
29.9	0\\
30	0\\
30.1	0\\
30.2	0\\
30.3	0\\
30.4	0\\
30.5	0\\
30.6	0\\
30.7	0\\
30.8	0\\
30.9	0\\
31	0\\
31.1	0\\
31.2	0\\
31.3	0\\
31.4	0\\
31.5	0\\
31.6	0\\
31.7	0\\
31.8	0\\
31.9	0\\
32	0\\
32.1	0\\
32.2	0\\
32.3	0\\
32.4	0\\
32.5	0\\
32.6	0\\
32.7	0\\
32.8	0\\
32.9	0\\
33	0\\
33.1	0\\
33.2	0\\
33.3	0\\
33.4	0\\
33.5	0\\
33.6	0\\
33.7	0\\
33.8	0\\
33.9	0\\
34	0\\
34.1	0\\
34.2	0\\
34.3	0\\
34.4	0\\
34.5	0\\
34.6	0\\
34.7	0\\
34.8	0\\
34.9	0\\
35	0\\
35.1	0\\
35.2	0\\
35.3	0\\
35.4	0\\
35.5	0\\
35.6	0\\
35.7	0\\
35.8	0\\
35.9	0\\
36	0\\
36.1	0\\
36.2	0\\
36.3	0\\
36.4	0\\
36.5	0\\
36.6	0\\
36.7	0\\
36.8	0\\
36.9	0\\
37	0\\
37.1	0\\
37.2	0\\
37.3	0\\
37.4	0\\
37.5	0\\
37.6	0\\
37.7	0\\
37.8	0\\
37.9	0\\
38	0\\
38.1	0\\
38.2	0\\
38.3	0\\
38.4	0\\
38.5	0\\
38.6	0\\
38.7	0\\
38.8	0\\
38.9	0\\
39	0\\
39.1	0\\
39.2	0\\
39.3	0\\
39.4	0\\
39.5	0\\
39.6	0\\
39.7	0\\
39.8	0\\
39.9	0\\
40	0\\
};
\addlegendentry{\tikzmatlabfontsscale CC}

\addplot [color=mycolor2, line width=1.0pt]
  table[row sep=crcr]{%
0	0\\
0.1	0\\
0.2	0\\
0.3	0\\
0.4	0\\
0.5	0\\
0.6	0\\
0.7	0\\
0.8	0\\
0.9	0\\
1	0\\
1.1	0\\
1.2	0\\
1.3	0\\
1.4	0\\
1.5	0\\
1.6	0\\
1.7	0\\
1.8	0\\
1.9	0\\
2	0\\
2.1	0\\
2.2	0\\
2.3	0\\
2.4	0\\
2.5	0\\
2.6	0\\
2.7	0\\
2.8	0\\
2.9	0\\
3	0\\
3.1	0\\
3.2	0\\
3.3	0\\
3.4	0\\
3.5	0\\
3.6	0\\
3.7	0\\
3.8	0\\
3.9	0\\
4	0\\
4.1	0\\
4.2	0\\
4.3	0\\
4.4	0\\
4.5	0\\
4.6	0\\
4.7	0\\
4.8	0\\
4.9	0\\
5	0\\
5.1	0\\
5.2	0\\
5.3	0\\
5.4	0\\
5.5	0\\
5.6	0\\
5.7	0\\
5.8	0\\
5.9	0\\
6	0\\
6.1	0\\
6.2	0\\
6.3	0\\
6.4	0\\
6.5	0\\
6.6	0\\
6.7	0\\
6.8	0\\
6.9	0\\
7	0\\
7.1	0\\
7.2	0\\
7.3	0.0177269282265926\\
7.4	0.0850632060500556\\
7.5	0.134865874036537\\
7.6	0.199465619573225\\
7.7	0.248547221827504\\
7.8	0.291375304701713\\
7.9	0.327881081266269\\
8	0.331987998546373\\
8.1	0.380938255369791\\
8.2	0.398374991502774\\
8.3	0.409349409454201\\
8.4	0.413908971687892\\
8.5	0.412315082559852\\
8.6	0.404577339803508\\
8.7	0.390808509274839\\
8.8	0.371199595238504\\
8.9	0.346949526759363\\
9	0.316454561393929\\
9.1	0.280691119983964\\
9.2	0.223396440961691\\
9.3	0.195590094820501\\
9.4	0.144805924908937\\
9.5	0.0910143492032264\\
9.6	0.0300803522323974\\
9.7	0\\
9.8	0\\
9.9	0\\
10	0\\
10.1	0\\
10.2	0\\
10.3	0\\
10.4	0\\
10.5	0\\
10.6	0\\
10.7	0\\
10.8	0\\
10.9	0\\
11	0\\
11.1	0\\
11.2	0\\
11.3	0\\
11.4	0\\
11.5	0\\
11.6	0\\
11.7	0\\
11.8	0\\
11.9	0\\
12	0\\
12.1	0\\
12.2	0\\
12.3	0\\
12.4	0\\
12.5	0\\
12.6	0\\
12.7	0\\
12.8	0\\
12.9	0\\
13	0\\
13.1	0\\
13.2	0\\
13.3	0\\
13.4	0\\
13.5	0\\
13.6	0\\
13.7	0\\
13.8	0\\
13.9	0\\
14	0\\
14.1	0\\
14.2	0\\
14.3	0\\
14.4	0\\
14.5	0\\
14.6	0\\
14.7	0\\
14.8	0\\
14.9	0\\
15	0\\
15.1	0\\
15.2	0\\
15.3	0\\
15.4	0\\
15.5	0\\
15.6	0\\
15.7	0\\
15.8	0\\
15.9	0\\
16	0\\
16.1	0\\
16.2	0\\
16.3	0\\
16.4	0\\
16.5	0\\
16.6	0\\
16.7	0\\
16.8	0\\
16.9	0\\
17	0\\
17.1	0\\
17.2	0\\
17.3	0\\
17.4	0\\
17.5	0\\
17.6	0\\
17.7	0\\
17.8	0\\
17.9	0\\
18	0\\
18.1	0\\
18.2	0\\
18.3	0\\
18.4	0\\
18.5	0\\
18.6	0\\
18.7	0\\
18.8	0\\
18.9	0\\
19	0\\
19.1	0\\
19.2	0\\
19.3	0\\
19.4	0\\
19.5	0\\
19.6	0\\
19.7	0\\
19.8	0\\
19.9	0\\
20	0\\
20.1	0\\
20.2	0\\
20.3	0\\
20.4	0\\
20.5	0\\
20.6	0\\
20.7	0\\
20.8	0\\
20.9	0\\
21	0\\
21.1	0\\
21.2	0\\
21.3	0\\
21.4	0\\
21.5	0\\
21.6	0\\
21.7	0\\
21.8	0\\
21.9	0\\
22	0\\
22.1	0\\
22.2	0\\
22.3	0\\
22.4	0\\
22.5	0\\
22.6	0\\
22.7	0\\
22.8	0\\
22.9	0\\
23	0\\
23.1	-0.0480377231972955\\
23.2	-0.107452797397068\\
23.3	-0.161679861253796\\
23.4	-0.210845140679398\\
23.5	-0.25498536076228\\
23.6	-0.294058787584519\\
23.7	-0.327959015183579\\
23.8	-0.35653073799881\\
23.9	-0.379586731265327\\
24	-0.396925169888841\\
24.1	-0.408346311533478\\
24.2	-0.413667507442416\\
24.3	-0.412735529050887\\
24.4	-0.405435336343281\\
24.5	-0.391694669509655\\
24.6	-0.371484200089887\\
24.7	-0.344813393959579\\
24.8	-0.311722668864393\\
24.9	-0.253538704022498\\
25	-0.228274187387398\\
25.1	-0.1765316535434\\
25.2	-0.118616051218581\\
25.3	-0.0545614483358185\\
25.4	-9.16519082563013e-05\\
25.5	0\\
25.6	0\\
25.7	0\\
25.8	0\\
25.9	0\\
26	0\\
26.1	0\\
26.2	0\\
26.3	0\\
26.4	0\\
26.5	0\\
26.6	0\\
26.7	0\\
26.8	0\\
26.9	0\\
27	0\\
27.1	0\\
27.2	0\\
27.3	0\\
27.4	0\\
27.5	0\\
27.6	0\\
27.7	0\\
27.8	0\\
27.9	0\\
28	0\\
28.1	0\\
28.2	0\\
28.3	0\\
28.4	0\\
28.5	0\\
28.6	0\\
28.7	0\\
28.8	0\\
28.9	0\\
29	0\\
29.1	0\\
29.2	0\\
29.3	0\\
29.4	0\\
29.5	0\\
29.6	0\\
29.7	0\\
29.8	0\\
29.9	0\\
30	0\\
30.1	0\\
30.2	0\\
30.3	0\\
30.4	0\\
30.5	0\\
30.6	0\\
30.7	0\\
30.8	0\\
30.9	0\\
31	0\\
31.1	0\\
31.2	0\\
31.3	0\\
31.4	0\\
31.5	0\\
31.6	0\\
31.7	0\\
31.8	0\\
31.9	0\\
32	0\\
32.1	0\\
32.2	0\\
32.3	0\\
32.4	0\\
32.5	0\\
32.6	0\\
32.7	0\\
32.8	0\\
32.9	0\\
33	0\\
33.1	0\\
33.2	0\\
33.3	0\\
33.4	0\\
33.5	0\\
33.6	0\\
33.7	0\\
33.8	0\\
33.9	0\\
34	0\\
34.1	0\\
34.2	0\\
34.3	0\\
34.4	0\\
34.5	0\\
34.6	0\\
34.7	0\\
34.8	0\\
34.9	0\\
35	0\\
35.1	0\\
35.2	0\\
35.3	0\\
35.4	0\\
35.5	0\\
35.6	0\\
35.7	0\\
35.8	0\\
35.9	0\\
36	0\\
36.1	0\\
36.2	0\\
36.3	0\\
36.4	0\\
36.5	0\\
36.6	0\\
36.7	0\\
36.8	0\\
36.9	0\\
37	0\\
37.1	0\\
37.2	0\\
37.3	0\\
37.4	0\\
37.5	0\\
37.6	0\\
37.7	0\\
37.8	0\\
37.9	0\\
38	0\\
38.1	0\\
38.2	0\\
38.3	0\\
38.4	0\\
38.5	0\\
38.6	0\\
38.7	0\\
38.8	0\\
38.9	0\\
39	0\\
39.1	0\\
39.2	0\\
39.3	0\\
39.4	0\\
39.5	0\\
39.6	0\\
39.7	0\\
39.8	0\\
39.9	0\\
40	0\\
};
\addlegendentry{\tikzmatlabfontsscale SC}

\end{axis}
\end{tikzpicture}%
 }
  \caption{Lateral acceleration when tracking a path without feedback, \textit{i.e.}, using only feedforward references. \label{fig:sim_lateral_acceleration}}
\end{figure}
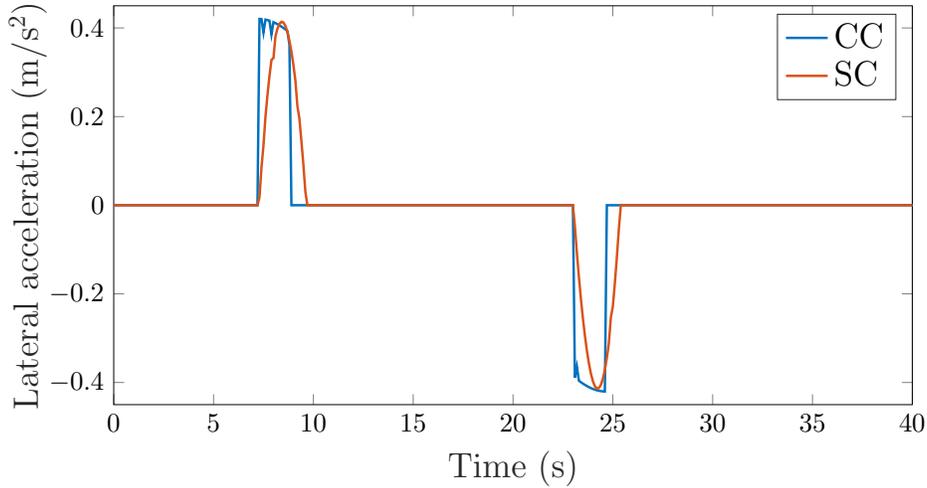

The lateral acceleration and jerk (acceleration rate) experienced by the vehicle are related to passenger comfort.
Figure~\ref{fig:sim_lateral_acceleration} shows the lateral accelerations for a vehicle following the reference steering profiles without feedback actuation.
It is seen that the CC path has large jerk values, which result from an aggressive steering actuation.
The SC path steering profile achieves smoother lateral acceleration profiles.

%%%%%%%%%%%%%%%%%
\section{Conclusions}
\label{sec:conclusions}

This paper presented the concept of SC paths.
SC paths respect not only the maximum steering angle constraints, but also maximum steering rate and acceleration constraints.
These properties ease the low-level controller task and introduce an higher degree of smoothness, improving the driving comfort and reducing actuator effort.
This is of importance when dealing with heavy-duty vehicles, which are characterized by slow actuator dynamics.

As future work, one could extend this approach so that the SC paths handle more cases besides that of a combination of two SC turns connected by a line segment. This would allow to connect configurations that lie close together.

Controllers could also be designed so that they take advantage of the smooth properties of the path, without requiring the computational burden of more complex control approaches that are design to withstand lower quality paths.

This work can also be extended into time optimal trajectory planning.
The velocity profile could be optimized, such that the vehicle performs the path in minimum time and obeys steering constraints.
Moreover, different curvature profiles could be optimized with respect to time, and abiding by the steering constraints.

%%%%%%%%%%%%%%%%%


\begin{thebibliography}{99}
%1
\bibitem{buehler2009darpa} M. Buehler \textit{et al.}, eds., The DARPA urban challenge: autonomous vehicles in city traffic. Vol. 56. springer, 2009.
%2
%\bibitem{choset2005principles} H. Choset \textit{et al.}, Principles of robot motion: theory, algorithms, and implementation. MIT press, 2005.
\bibitem{lavalle2006planning} S. LaValle, Planning algorithms. Cambridge University Press press, 2006.
%3
\bibitem{reuter1998} J. Reuter, "Mobile robots trajectories with continuously differentiable curvature: an optimal control approach" Proceedings of the 1998 IEEE/RSJ International Conference on Intelligent Robots and Systems (1998): 38-43.
%4
\bibitem{nagy2001trajectory} B. Nagy and A. Kelly, "Trajectory generation for car-like robots using cubic curvature polynomials" Field and Service Robots 11 (2001).
%5
\bibitem{MCCRAE2009452} J. McCrae and K. Singh, "Sketching piecewise clothoid curves" Computers \& Graphics 33, no. 4 (2009): 452-461.
%6
\bibitem{bianco2004smooth} A. Piazzi \textit{et al.}, "$\eta^3$-Splines for the Smooth Path Generation of Wheeled Mobile Robots" IEEE Transactions on Robotics 23, no. 5 (2007): 1089-1095.
%7

\bibitem{kuwata2009real} Y. Kuwata \textit{et al.}, "Real-time motion planning with applications to autonomous urban driving" IEEE Transactions on Control Systems Technology 17, no. 5 (2009): 1105-1118.

\bibitem{karaman2011anytime} S. Karaman \textit{et al.}, "Anytime motion planning using the RRT*," in Robotics and Automation (ICRA), 2011 IEEE Int. Conf. on, pp. 1478-1483. IEEE, 2011.

\bibitem{shima2006multiple} T. Shima \textit{et al.}, "Multiple task assignments for cooperating uninhabited aerial vehicles using genetic algorithms" Computers \& Operations Research 33, no. 11 (2006): 3252-3269.

\bibitem{dolgov2010path} D. Dolgov \textit{et al.}, "Path planning for autonomous vehicles in unknown semi-structured environments" The International Journal of Robotics Research 29, no. 5 (2010): 485-501.

\bibitem{dubins1957curves} L. E. Dubins, "On curves of minimal length with a constraint on average curvature, and with prescribed initial and terminal positions and tangents" American Journal of mathematics 79, no. 3 (1957): 497-516.

\bibitem{reeds1990optimal} J. Reeds and L. Shepp, "Optimal paths for a car that goes both forwards and backwards" Pacific journal of mathematics 145, no. 2 (1990): 367-393.

\bibitem{fraichard2004reeds} T. Fraichard and A. Scheuer, "From Reeds and Shepp's to continuous-curvature paths" IEEE Transactions on Robotics 20, no. 6 (2004): 1025-1035.

\bibitem{lima2015clothoid} P. F. Lima \textit{et al.}, "Clothoid-based model predictive control for autonomous driving" In Control Conference (ECC), 2015 European, pp. 2983-2990. IEEE, 2015.

\bibitem{parlangeli2010dubins} G. Parlangeli and G. Indiveri, "Dubins inspired 2D smooth paths with bounded curvature and curvature derivative" IFAC Proceedings Volumes 43, no. 16 (2010): 252-257.

\bibitem{videogames} M. Cirillo, "From videogames to autonomous trucks: A new algorithm for lattice-based motion planning" 2017 IEEE Intelligent Vehicles Symposium (IV), pp. 148-153.

\end{thebibliography}
\end{document}